\documentclass[fleqn,usenatbib]{mnras}

\usepackage[T1]{fontenc}

\DeclareRobustCommand{\VAN}[3]{#2}
\let\VANthebibliography\thebibliography
\def\thebibliography{\DeclareRobustCommand{\VAN}[3]{##3}\VANthebibliography}

\usepackage{graphicx}	
\usepackage{amsmath}	
\usepackage{amssymb}	
\usepackage{multirow}
\usepackage{booktabs, tabularx}
\usepackage{listings}

\usepackage{algorithmicx}
\usepackage{algorithm}
\usepackage{algpseudocode}

\title[The pulsar tree]{
Visualizing the pulsar population using graph theory}

\author[García, Torres, \& Patruno]{
C. R. Garc\'{i}a$^{1,3}$\thanks{E-mail: crodriguez@ice.csic.es}, 
Diego F. Torres$^{1,2,3}$\thanks{E-mail: dtorres@ice.csic.es}, 
Alessandro Patruno$^{1,3}$\thanks{E-mail: patruno@ice.csic.es}
\\
$^{1}$Institute of Space Sciences (ICE, CSIC), Campus UAB, Carrer de Can Magrans s/n, 08193 Barcelona, Spain\\
$^{2}$Institució Catalana de Recerca i Estudis Avançats (ICREA), E-08010 Barcelona, Spain \\
$^{3}$Institut d’Estudis Espacials de Catalunya (IEEC), 08034 Barcelona, Spain
}

\pubyear{2022}

\usepackage{newtxtext,newtxmath}

\begin{document}
\label{firstpage}
\pagerange{\pageref{firstpage}--\pageref{lastpage}}
\maketitle

\begin{abstract}
The $P\dot{P}$ diagram is a cornerstone of pulsar research. It 
is used in multiple ways for classifying the population, understanding evolutionary tracks, identifying issues in our theoretical reach, and more.
However, we have been looking at the same plot for more than five decades. A fresh appraisal may be healthy. Is the $P\dot{P}$-diagram the most useful or complete way to visualize the pulsars we know? Here we pose a fresh look at the information we have on the pulsar population. First, we use  principal
components analysis over magnitudes depending on the intrinsic pulsar’s timing properties (proxies to relevant physical pulsar features), to analyze whether the information contained by the pulsar’s period and period derivative is enough to describe the variety of the pulsar population. Even when the variables of interest depend on $P$ and $\dot{P}$, we show that $P\dot{P}$ are not principal components. Thus, any distance ranking or visualization based only on $P$ and $\dot{P}$ is potentially misleading. Next, we define and compute a properly normalized distance to measure pulsar nearness, calculate the minimum spanning tree of the population, and discuss possible applications. The pulsar tree hosts information about pulsar similarities that go beyond $P$ and $\dot{P}$, and are thus naturally difficult to read from the $P\dot{P}$-diagram. We use this work to introduce the pulsar tree website ({\url{http://www.pulsartree.ice.csic.es}}),
containing visualization tools and data to allow users to gather information in terms of MST and distance ranking.

\end{abstract}


\section{Introduction}
\label{intro}

Ever since the discovery of the first pulsar \citep{CP1919}, the plot containing the period derivative versus the period of every pulsar found, or simply, the $P \dot P$-diagram, has been 
used as a way of summarizing our knowledge and guiding our research on the pulsar population. Classes of pulsars and possible links among them are referred to in this plot, as it is what we know about their possible evolutionary tracks along the pulsar's lifetime (see e.g., \cite{Enoto2019} for a review). However, we have been looking at the same plot for more than five decades. A fresh appraisal may be healthy. Is the $P \dot P$-diagram the most useful or complete way to visualize the pulsars we know? Does it introduce any unwarranted bias on what we consider to be similar pulsars? Here, we use principal components analysis (PCA, e.g., \cite{Pearson1901,Shlens2014}) to show that even when the variables considered to describe a pulsar would all depend on $P$ and $ \dot P$, the variance of the population is not contained in the variance of the latter quantities. Thus, we note that any nearness ranking or visualization based only on $P$ and $\dot P$ is potentially misleading. Next we define and compute a properly normalized distance to measure nearness from one pulsar to another, calculate the minimum spanning tree (MST, e.g., \cite{Gower1969}) of the pulsar population, and discuss possible applications.
The pulsar tree hosts information about pulsar similarities that go beyond $P$ and $\dot P$, and are thus naturally difficult to read from the $P\dot P$-diagram. We also introduce here an online tool encompassing all our results, as well as allowing a user to focus on user-defined problems.

The MST is a graph that connects points in a multi-dimensional space. Each point (or node) is linked to at least another by an edge, whose length is associated with a given distance. 
The edges of an MST are chosen so that the the sum of their lengths is minimal and all nodes are linked, implying no cycles are present (no paths starting and returning to the same node).
Graph theory shows that as long as distances are distinct, the MST is unique. Its very definition intuitively implies that the MST is an optimization technique. In fact, it was widely used in engineering problems, starting from their original application developed by Borůvka in 1926, for the distribution of electricity in Moravia, see \cite{Nesetril2021}. Currently, MSTs are used from analyzing cognitive impairment \citep{Simon2021} to risk in financial markets \citep{Pozzi2013}.
Early usage in scientific problems include describing the interrelationship of species or genetics (see the work of Florik in the 1950's and Edwards in 1960's as commented in \cite{Hartigan1981} and \citep{Winther2018}, respectively), disciplines in which it is a widespread technique.
In astronomy, it has been used for finding high-energy sources \citep{Campana2013}, establishing differences between cluster and field stars \citep{Sanchez2018}, detecting filaments \citep{Pereyra2020}, galaxy clustering \citep{Barrow1985}, and cosmology \citep{Naidoo2020}.
It has also been used in the analysis of event samples in particle colliders \citep{Rainbolt2017}, or cosmic-rays \citep{Harari2006}. This non-exhaustive reference list is just an example of growing interest in MST use across different fields. Despite this interest, it has been barely been used in relation to pulsars. To our knowledge there is only one related publication \citep{Maritz2016} using 11 handpicked objects. The aim was to show that an MST could distinguish binaries from isolated pulsars using the dispersion measure as distance. In this paper, we explore using the MST to provide a novel classification, alerting, and visualization tool for pulsars. 

\section{The pulsar variance}
\label{subsec:variance}

\subsection{Variables definition}
\label{subsec:variables}

We consider pulsars listed in the ATNF Catalog \citep{ATNF-Catalog}, including radio
pulsars, X-ray and/or gamma-ray pulsars, and magnetars for which coherent pulsations have been detected. Accretion-powered pulsars such as, e.g., SAX J1808.4-3658, are not included in this table, however. The current number of pulsars listed in the catalog is 3282, of which 2509 have a known period and period derivative (larger than 0). From the latter, 2242 are isolated pulsars and 267 are pulsars residing in binary systems. All the methods considered in this work will be applied to this set as a whole, without making any distinction among the pulsars in it.
For characterizing the pulsar population, and ultimately defining a `distance' from one pulsar to another, we shall consider the following physical set of pulsar variables (see, e.g., \cite{Lorimer2012} for reference):

\begin{itemize}
  \item Spin period:\\ $P$ [s],
  \item Spin period derivative:\\ $\dot P$ [s s$^{-1}$],
  \item Surface magnetic flux density (equator): \\ $B_{s} = (3c^{3}I)^{1/2}/(8\pi^{2} R^{6} \sin^{2}\alpha)^{1/2}
  \sqrt{P\dot P}  \simeq 3.2 \times 10^{19}  P^{1/2}\dot{P}^{1/2} {\rm G}$,
  \item Magnetic field at the light cylinder:\\ $B_{lc}=B_{s}(\Omega R)^{3}/c^{3} \simeq 3\times 10^8 P^{-5/2} \dot{P}^{1/2} {\rm G}$,
  \item Spin-down energy loss rate:\\ $\dot{E}_{sd} = 4\pi ^{2}I\dot P P^{-3}\simeq  3.95 \times 10^{46} P^{-3}  \dot{P} \;\;\; {\rm erg} \; {\rm s^{-1}}$,
  \item Characteristic age: \\ $\tau_c = {{P}}/{2\dot P} \simeq 15.8 \times P \dot P^{-1} {\rm Myr}$, 
  \item Surface electric voltage: \\ $\Delta \Phi = ({B_s 4\pi^2 R^3})/({2 c P^2}) \simeq
  6.3 \times 10^{5} P^{-3/2} \dot{P}^{1/2} {\rm V}$,
  \item Goldreich-Julian charge density: \\ $\eta_{GJ} = (\Omega B_{s})/(2\pi ce)\simeq  7 \times 10^{10}P^{-1/2}  \dot{P}^{1/2} {\rm cm^{-3}}$.
\end{itemize}

Here the moment of inertia $I$ was assumed as $10^{45}$ g cm$^{2}$, the radius of the star $R$ was assumed as $10$ km, and the inclination $\alpha$ between the magnetic and rotation axes as $90^o$.
The remaining constants ($c, e$) have the usual meaning.

The measurable quantities $P$ and $\dot P$ are the leading magnitudes in this set of variables, from which all others are calculated using the rotating dipole model, as is usual for pulsar estimations. 

The surface magnetic field and spin-down power are basic magnitudes critical to characterize the pulsars' energetics and their magnetospheres. In our set, they are complemented with others to incorporate the fact that dissimilar pulsars (e.g., millisecond and normal) can have similar magnetospheres. This is partly described by the magnetic field at the light cylinder, which is similar in both cases. The voltage gives the potential drop between the magnetic pole and the edge of the polar cap. It is thought here to represent the variety introduced by the electromagnetic configuration, for which another parameter of interest is the Goldreich-Julian charge density, $\propto B_s / P$. Note that these magnitudes, being all functions of $P$ and $\dot P$, can have a relationship between themselves, as just noted.

The mass and the radius of the neutron stars would ideally also be considered as part of the variables of interest in our study. However, this information is only available for a very small percentage of the sample.

Other variables of interest are those related to the birth properties of pulsars, such as the initial spin-down power, the initial magnetic field, or the spin-down timescale. However, these are not known for most pulsars in our sample. They all depend on the unknown (except for a few) pulsars' real age (for which the characteristic age $\tau_c$ is only a proxy). Similarly, the braking index is measured for just a handful of pulsars (and in addition, it is known that it may vary significantly). Estimates from it using $\ddot P$ would significantly reduce the sample. Finally, other measurable quantities exist that are unrelated to intrinsic properties (transverse velocities, DM, distances) and/or are known for a limited number of objects. Using luminosities and other properties at different frequencies (e.g., fluxes, pulse shapes, peak separation, etc) would also significantly cut the sample, would be affected by extrinsic conditions (absorption, distance), and/or would incorporate parameters that are difficult to compare for the population as a whole. These are left for analysis in future work, where particular sub-samples will be looked at in more detail taking into account different variables, focusing here only on intrinsic pulsar features to introduce the technique.

\subsection{Treating variables}
\begin{figure*}
  \includegraphics[width=1\textwidth]{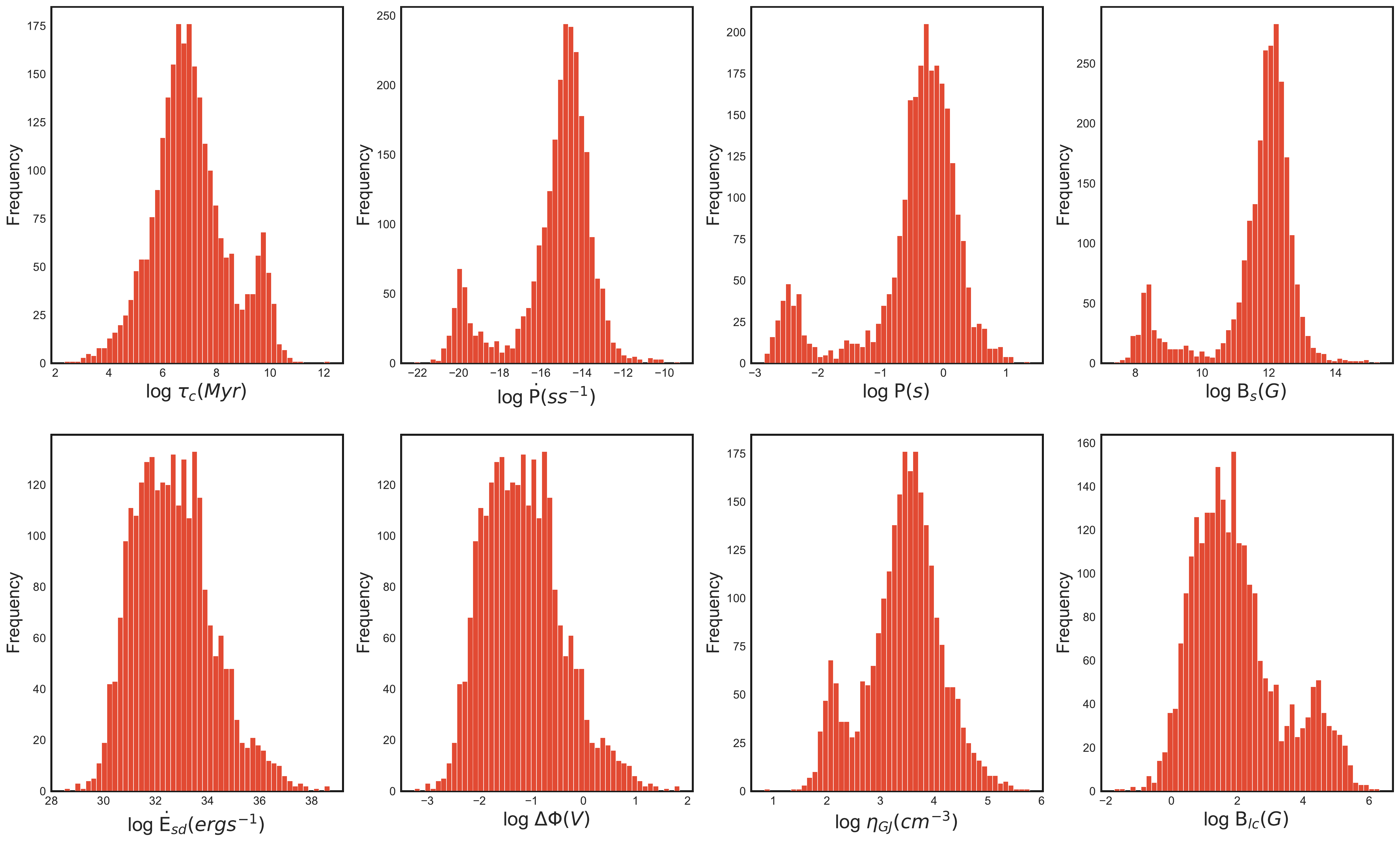}
  \centering
  \caption{Distribution of the logarithm of the 8 variables used for the complete set of 2509 pulsars. }
  \label{figure1}
\end{figure*}

The values of the magnitudes considered may differ by several orders of magnitude for different pulsars. The distributions of the logarithm of these variables are shown in Fig. \ref{figure1}.
The distributions are not normal (as is also the case for the set of original variables, without logarithm). Two clear populations of millisecond and normal pulsars appear separately in all plots, except the two related to the spin-down power and voltage. As is obvious from their centralization (mean and median) and dispersion (standard deviation and interquartile range (IQR)), the log-variables are orders of magnitude closer together. Given that the distributions are not normal, we use the {\it robust scaler} for normalizing the log variables,
\begin{eqnarray}
x_{i}^\dag = \frac{x_{i}-Q_{2}}{IQR}
~,\label{eq:robust_scaler.}
\end{eqnarray}
where  the $\dag$-symbol represents that the quantity $x_{i}$ has been normalized, $Q_{1}$, $Q_{2}$ and $Q_{3}$ represent the 1st quartile, median, and 3rd quartile of the distribution, respectively, and IQR is the interquartile range, $(Q_{3}-Q_{1})$. It can be seen that the distributions of the variables after being normalized have a median equal to zero and an IQR equal to one. Note that if we take the logarithm of the variables and then apply Eq. (\ref{eq:robust_scaler.}) to normalize it, relations between variables are more clearly uncovered. For instance, $\dot{E}_{sd}$ and $\Delta \Phi$, lead to $\log \dot{E}_{sd}^\dag = \log \Delta \Phi^\dag$, that is also visible in the corresponding distributions of Fig. \ref{figure1}. Considering both of them at once in defining the nearness of two given pulsars relates to the fact that the physical meaning represented by the two original magnitudes is different.

\section{Principal components analysis}
\label{subsec:pca-pulsars}

\begin{figure*}
  \centering
  \includegraphics[width=1\textwidth]{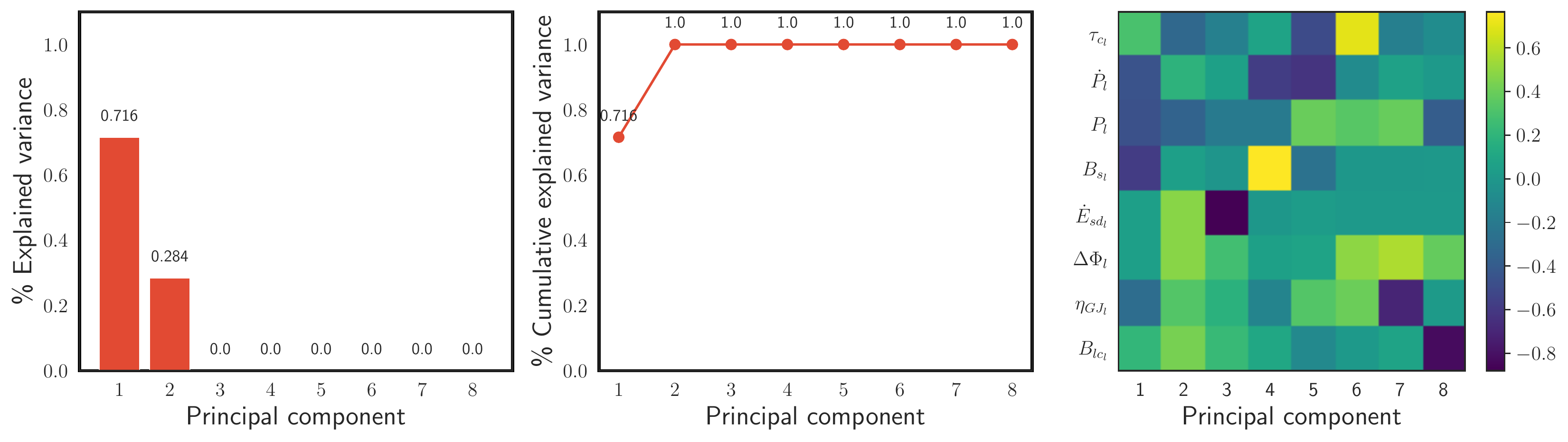}
  \caption{
  PCA results using the logarithm of the set of variables for the whole population of 2509 pulsars.
  The left panel shows the explained variance contributed by each of the PCs according to the eigenvalues of the covariance matrix. It represents the amount of information contained in each PC.
  The central panel shows the cumulative variance explained by the new set of variables which has been defined through the PCA analysis. The right panel shows the 'weight' that each variable has with respect to each PC. This value is the coefficient that is held in each eigenvector. Negative values imply that the variable and the PC in question are negatively correlated. Conversely, a positive value shows a positive correlation between the PC and the variable. 
  }
  \label{figure2}
\end{figure*}

PCA is especially suitable to isolate the main factors introducing variance in a population when the variables known for it are not independent (see Appendix 1). Since 6 of the variables taken to describe the intrinsic properties of the pulsar population are in fact computed from $P$ and $\dot P$, one can intuitively conclude that 2 principal components (PCs) are needed to describe our population variance. However, the latter {\it is not} contained in the variance of $P$ and $\dot P$. Said otherwise, $P$ and $\dot P$ are not the two principal components needed. Thinking only in terms of $P$ and $\dot P$ to compare pulsars may thus be misleading, except in the extreme case where these values are simply the same. Fig. \ref{figure2} shows the result of the PCA analysis. 
The two PCs are, 
\begin{eqnarray}
    PC_1 &=& 0.21 B_{{lc}_l}^\dag  - 0.29 \eta_{{GJ}_l}^\dag  + 0.05 \Delta \Phi_l^\dag  + 0.05 \dot{E}_{{sd}_l}^\dag  -0.46 \dot P_l^\dag \nonumber \\
    && -0.59 B_{s_l}^\dag -0.47 P_l^\dag + 0.29 \tau_{c_l}^\dag,  
    \nonumber \\
    && 
    {
    {\rm (71.6\% \;\; of 
    \;\;the 
    \;\;explained
    \;\;variance)}},
    \label{eq:PC1_8V}\\
    PC_2 &=& 0.43 B_{{lc}_l}^\dag  + 0.32 \eta_{{GJ}_l}^\dag  + 0.47 \Delta \Phi_l^\dag  + 0.47 \dot{E}_{{sd}_l}^\dag  +0.19 \dot P_l^\dag \nonumber \\
    && + 0.05 B_{s_l}^\dag -0.36 P_l^\dag - 0.32 \tau_{c_l}^\dag, 
    \nonumber \\
    && 
    {
    {\rm (28.4\% \;\; of 
    \;\;the 
    \;\;explained
    \;\;variance)}},
    \label{eq:PC2_8V}
\end{eqnarray}
where the $^\dag$-quantities refer to the normalized ones as in Eq. (\ref{eq:robust_scaler.}), and the sub-index $l$ stands to note that the normalization is applied to the logarithm of the variable.
After some algebra (see Appendix 1) these latter equations can be reformulated as
\begin{eqnarray}
 PC_1 &=& -8.471 - 1.178 \log P - 0.832 \log \dot{P}, \label{eq:PC1_2V}\\
 PC_2 &=& 14.182 - 2.931 \log P + 1.105 \log \dot{P} 
    .\label{eq:PC2_2V}
\end{eqnarray}
Note that no dag marking is herein needed, as we have absorbed the corresponding IQR and median of each variable into the coefficients and that the units of $P$ and $\dot P$ are as in Eq. (\ref{eq:PC2_8V}).

\begin{figure*}
  \includegraphics[width=1\textwidth]{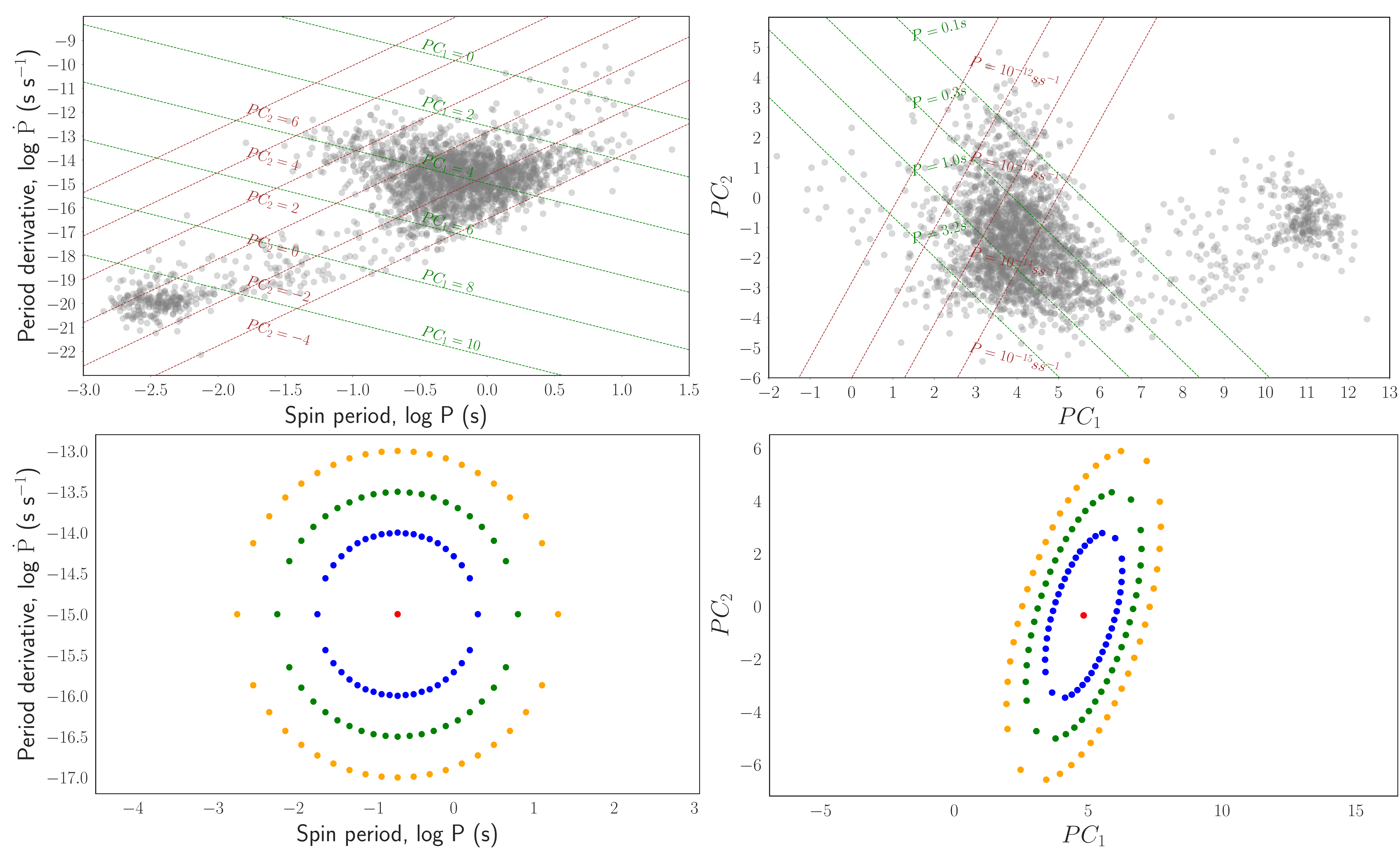}
  \centering
\caption{Top-Left panel: Representation of the total pulsar population as a function of the $P$ and $\dot P$ variables. Constant values of $PC_{1}$ and $PC_{2}$ are shown in green and brown lines, respectively, according to Eqs. (\ref{eq:PC1_2V}) and (\ref{eq:PC2_2V}). Top-right panel: Representation of the total pulsar population as a function of $PC_{1}$ and $PC_{2}$ whose values are obtained according to the 8 variables taken into account as shown in Eqs. (\ref{eq:PC1_8V}) and  (\ref{eq:PC2_8V}). Constant values of $P$ and $\dot P$ are shown in green and brown lines, respectively, taking into account the Eqs. (\ref{eq:PC1_2V}) and (\ref{eq:PC2_2V}).Bottom-left panel: Synthetic pulsars positioned circularly at different radii from a given centre. Bottom-right panel: Transformation of the circle through the Eqs. (\ref{eq:PC1_2V}) and (\ref{eq:PC2_2V}).}
       \label{figure3}
\end{figure*}

The left panel of Fig. \ref{figure3} shows the pulsar population in the $\log P - \log \dot P$ together with lines representing equal values of the principal components. The right panel of Fig. \ref{figure3} shows the same pulsars but directly in the plane $PC_1, PC_2$. Nearness in one plane has not the same meaning as it has in the other. To exemplify this, we plot a circle in the $P,\dot P$ diagram and see how the circle transforms to the $PC_1,PC_2$ plane via the equations above. This is shown in the second row of Fig. \ref{figure3}; the difference in relative distances can be up to a factor of 3 or more. This change of shape advances the idea that any sort of nearness ranking will be affected if considering the PCs, instead of $P$ and $\dot P$ (further comments about this can be found in the Appendix). 

\section{minimum spanning tree of the pulsar population}
\label{subsec:mst-pulsars}

\begin{figure*}
\centering
\includegraphics[width=\textwidth]{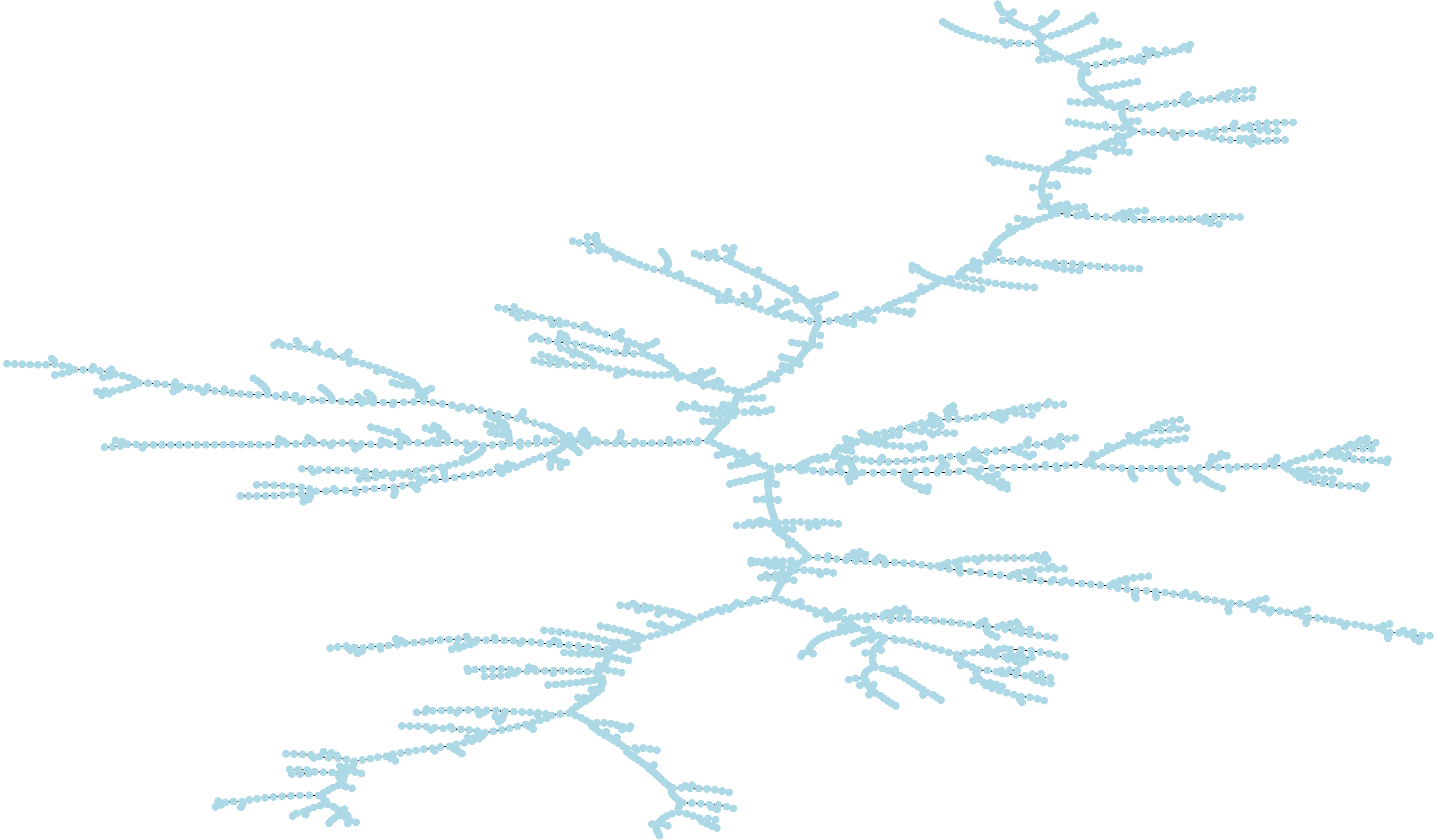}
\caption{A different look at the pulsar population. MST-graph (2509, 2508) based on the complete, undirected and weighted graph $G(2509,3146286)$ for the 2509 pulsars and their full combination of weights computed from their Euclidean distance among 8 normalized variables (or the equivalent 2 PCs). Each node in the MST represent a pulsar. Branches group pulsars with particular characteristics.}
       \label{figure4}
\end{figure*}

Appendix 2 introduces all concepts needed to understand and compute an MST, and we take this for granted in what follows. We define an Euclidean distance using the eight normalized (via Eq. \ref{eq:robust_scaler.}) logarithm of the variables introduced above. Equivalently, after our analysis of the previous section, we can use just two variables according to the PCA analysis, that contains the whole population variance. Both choices end up producing the same MST (and thus the results from the analysis that follows from it are the same) but the latter is less demanding given the reduced dimensionality of the problem. With this Euclidean distance, we first obtain a complete, undirected and weighted graph $G(V, E)=G(2509,3146286)$, with $|V|$ nodes and $|E|$ edges, where each edge is defined by a specific weight value. From that, we obtain the pulsar MST, which is shown in Fig. \ref{figure4}. 

\subsection{Branch analysis and pulsar classification in the MST}
\label{subsec:branching}

\begin{figure*}
  \includegraphics[width=1\textwidth]{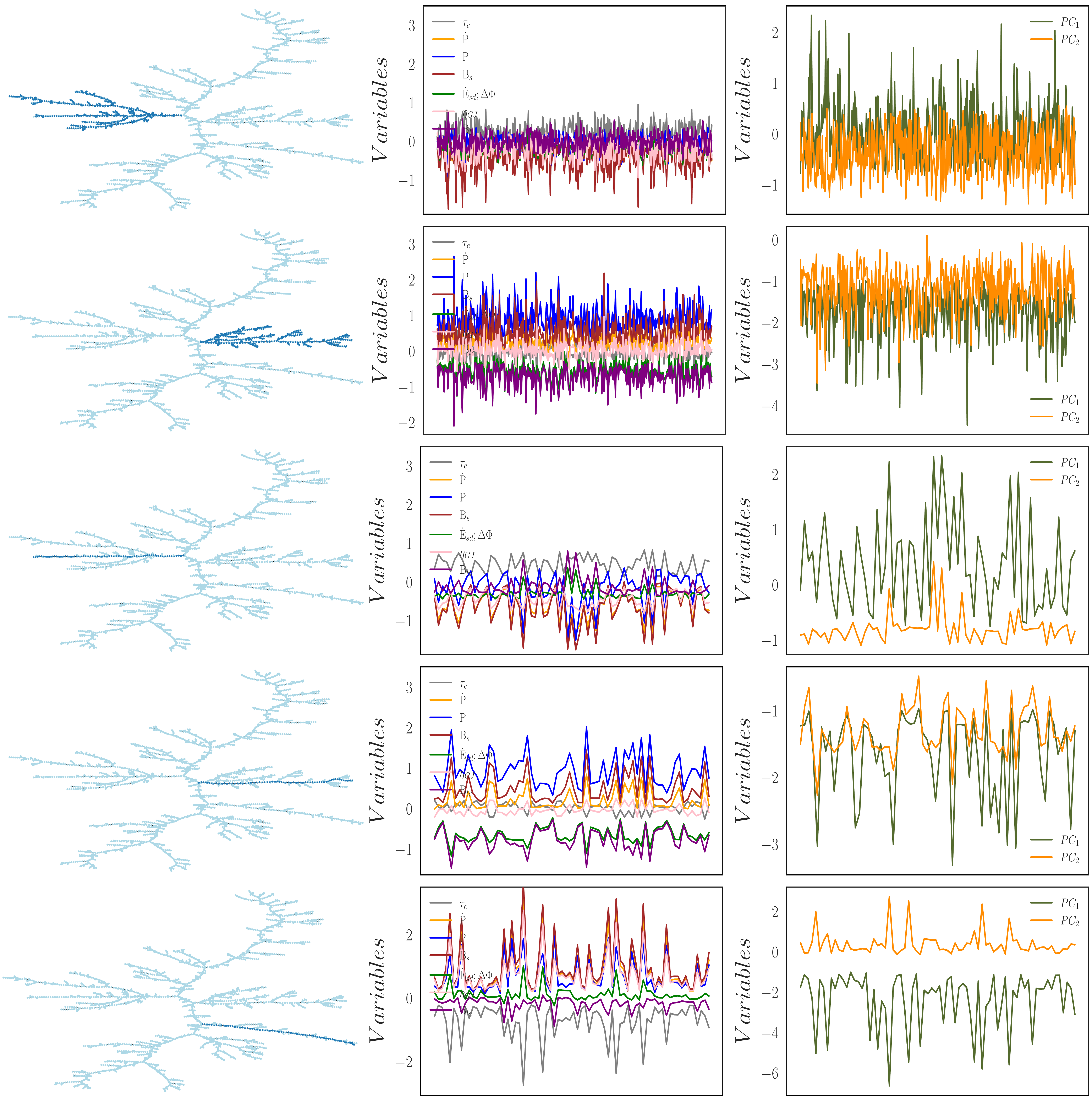}
  \centering
\caption{Top two rows: In the left, nodes from different branches are highlighted in dark blue in the MST. The middle and right panels show the behavior of the 8 normalized variables and of the 2 principal components when following an arbitrary mixing of the nodes in the highlighted part of the MST. Third, fourth and fifth rows: the same, but now arbitrarily mixing the nodes of only one branch. Note that for the purpose of plotting both $PC_1$ and $PC_2$ in the same scale in the right panels of Fig. \ref{figure5}, and also Fig. \ref{figure6}, we have subtracted the corresponding mean -from the whole set- from each set of values.
}
       \label{figure5}
\end{figure*}

\begin{figure*}
  \includegraphics[width=1\textwidth]{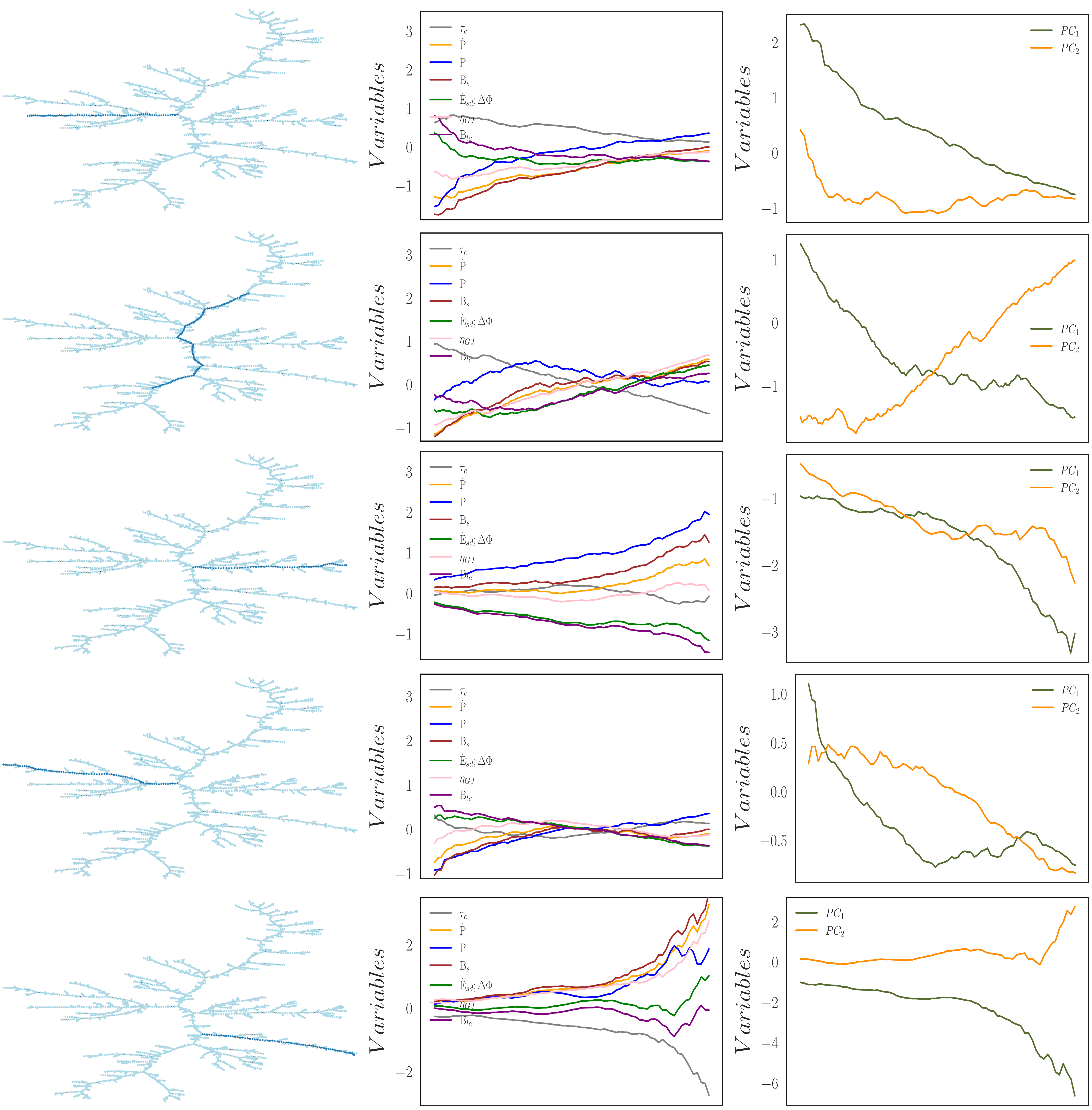}
  \centering
\caption{Similar to the panels in Fig. \ref{figure5}, but here following the node sequence as appearing along the marked branch of the MST. The node sequence is here read from left to right (except in those branches that appear mostly vertical, where they are read from top to bottom).
}
       \label{figure6}
\end{figure*}

\begin{figure*}
  \includegraphics[width=1\textwidth]{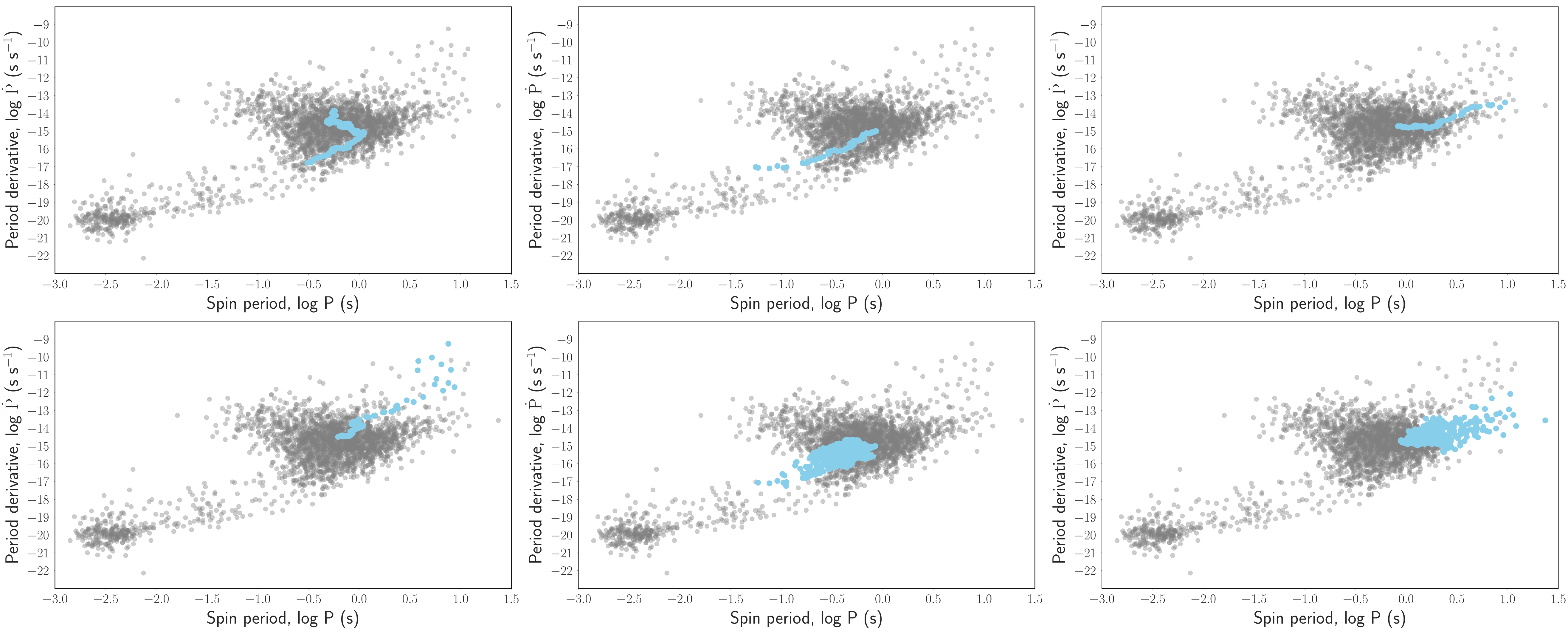}
  \centering
\caption{Positions of the branches analyzed in Figs. 
\ref{figure5} and 
\ref{figure6}. We show the main trunk of the tree (top left) followed by the branch depicted in 
row 3 of Fig. \ref{figure5} (or row 1 of Fig. \ref{figure6}), row 4 of Fig. \ref{figure5} (or row 3 of Fig. \ref{figure6}), and row 5 of Fig. \ref{figure5} (or row 5 of Fig. \ref{figure6}). the last two panels show the corresponding pulsars to the branches marked in rows 1 and 2 of Fig. \ref{figure5}.
}
       \label{figure7}
\end{figure*}

As shown in Fig. \ref{figure5}, mixing nodes from different branches of the MST produces a scattered distribution of variables.  
This is a generic behavior that happens for any mixing of the branches in the MST, and for any mixing of the nodes even within a single branch (see the panels in the three last rows of Fig. \ref{figure5}). If we read the MST in a disordered manner, nothing is learned from it. Instead, Fig. \ref{figure6} shows that if we  choose one of the branches at a time and run along with the nodes in it in an orderly manner, a smooth behavior of the variables naturally appears. Mixing branches of the pulsar tree is equivalent to grouping pulsars by their nearness in the $P \dot P$-diagram.
This is visually shown in Fig. \ref{figure7}. The pulsar tree hosts information that is difficult to read from the $P\dot P$-diagram.

\subsubsection{The MST as a descriptive tool}
\begin{figure*}
  \includegraphics[width=1\textwidth]{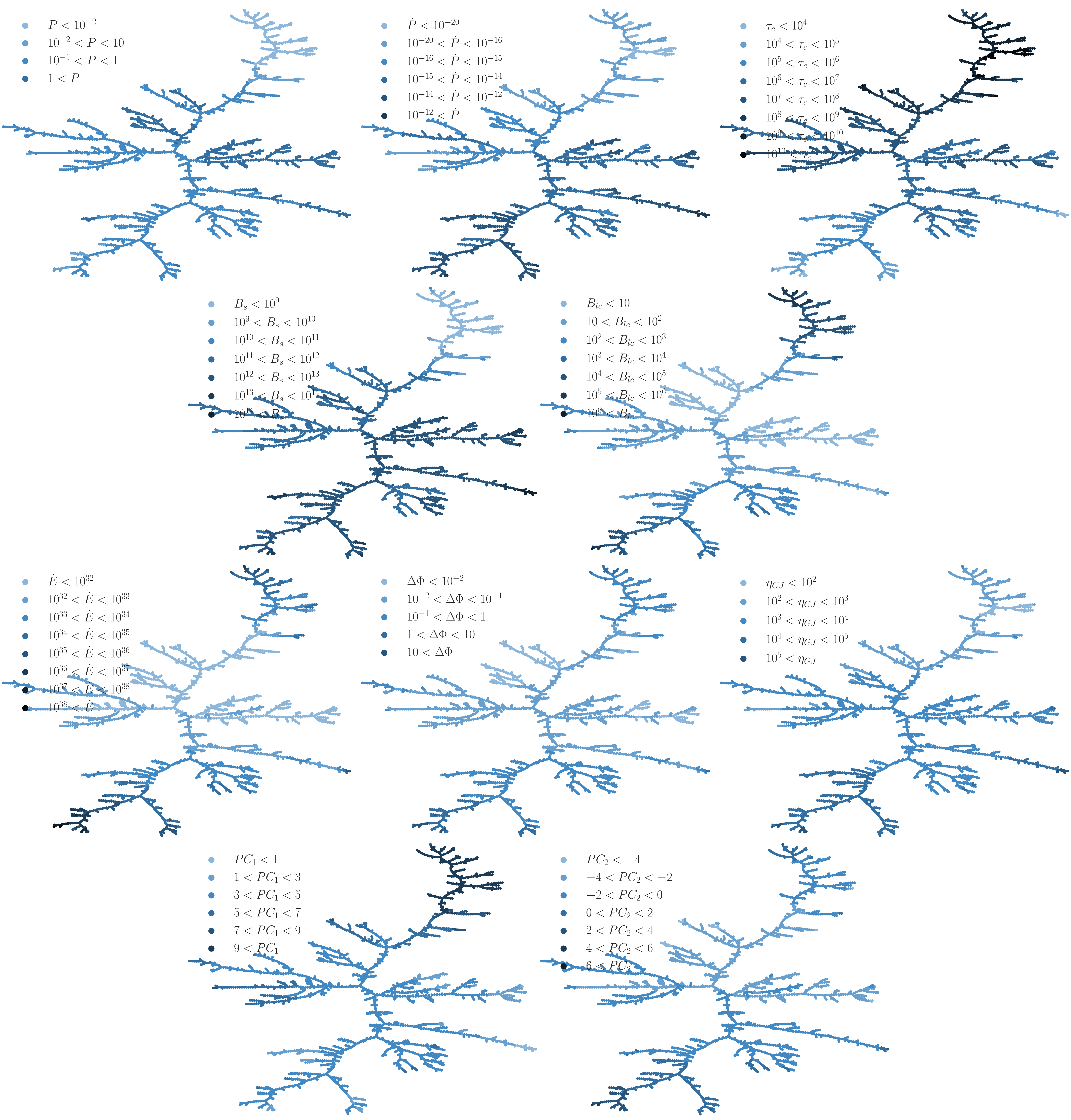}
  \centering
\caption{
Representation of the values of the different pulsar magnitudes considered directly onto the MST. The first row shows $P$, $\dot P$ and $\tau$; the second row shows the magnetic field at the surface and at the light cylinder; the third row shows the spin-down power, the voltage and the Goldreich-Julian current, and finally the last row shows directly 
the principal components $PC_{1}$ and $PC_{2}$ values.}
       \label{figure8}
\end{figure*}

The ordering introduced by each of the branches is indicative that there is an internal physical grouping in the MST. This is shown in Fig. \ref{figure8}, where we also show the variation of the principal components $PC_1$ and $PC_2$. These variations serve to visualize the physical properties of different pulsar classes, and understanding them may lead to physical connections among pulsars, or links in their evolution. To emphasize this, we shall observe how some known groups of pulsars locate in the MST.

\begin{figure*}
  \includegraphics[width=1\textwidth]{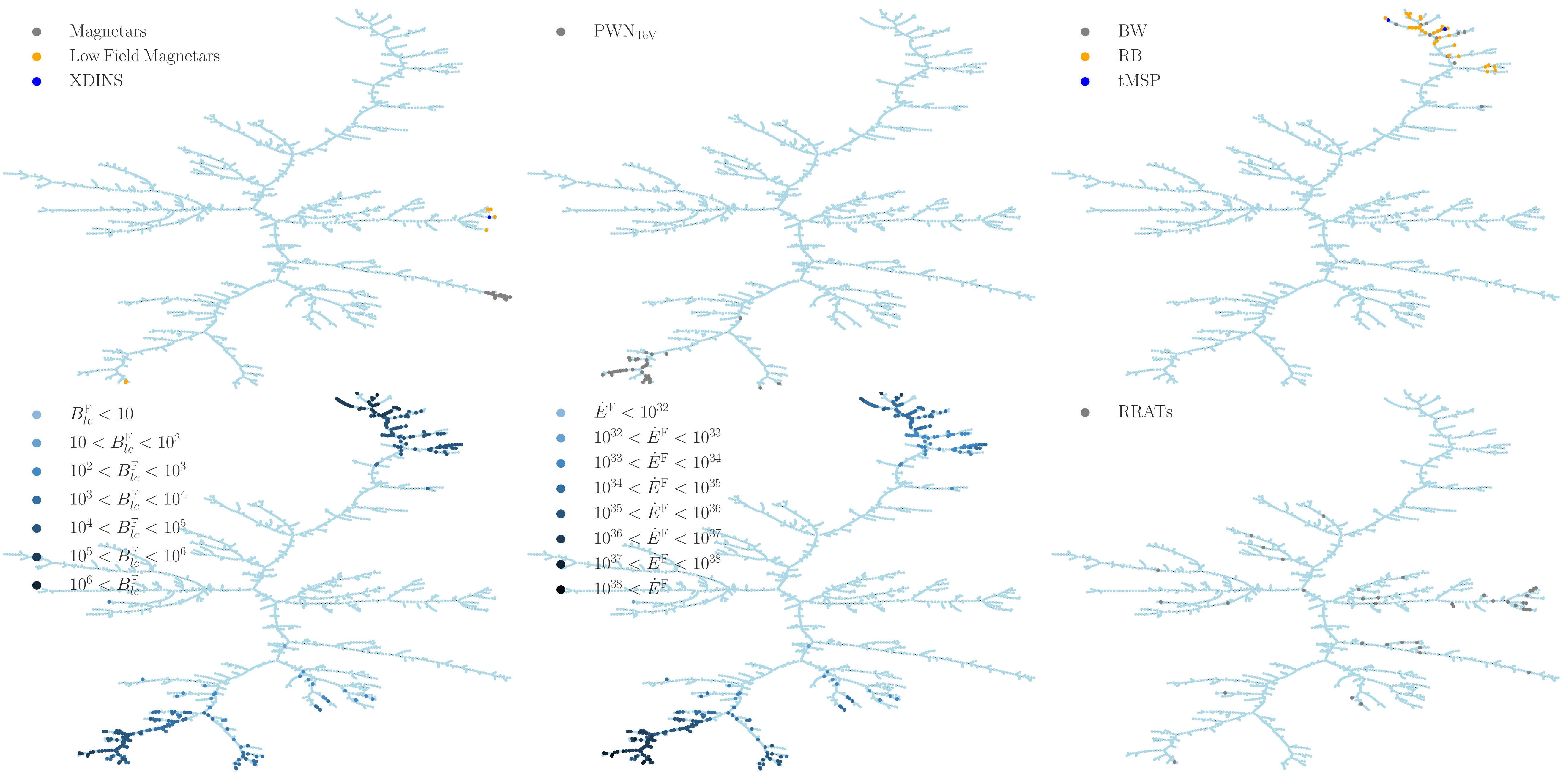}
  \centering
\caption{
Representation of the members of different pulsar classes onto the $(PC_1,PC_2)$ MST. The first row shows magnetars; redbacks, black widows and transitional pulsars, and pulsars associated with TeV pulsar wind nebulae. The first two panels second row shows the {\it Fermi}-LAT pulsars (thus the superscript F), further characterized against their spin-down power and light cylinder radius, as depicted in Fig. \ref{figure8}. The last panel
shows the RRATs. 
}
       \label{figure9}
\end{figure*}

The main tree trunk travels from young and energetic pulsars in the bottom\footnote{In what follows, we use `bottom', `top', `right', or `left' to facilitate the referral to a particular position in the MST shown. However, we emphasize that (in lack of axis) what matters is not the representation but the edges among nodes --or, in technical terms, the adjacency matrix. } to millisecond pulsars in the top (see panels -referred from left to right and top to bottom- 1, 2, and 5 of Fig. \ref{figure8}). However, other than representing the overall separation of the binary pulsars from the rest of the sample, the variables used in this MST do not allow us to dig deeper into sub-population of binaries: they are heavily affected by the evolutionary processes that occur for MSPs during the accreting (recycling) phase. As an example, consider black-widows (BWs), redbacks (RBs), and transitional pulsars (tMSPs) (see \cite{Papitto2022} for reviews). Black Widows and redbacks are binary MSPs where the companion is a semi-degenerate star or a main sequence star, respectively. The tMSPs (of which only two tMSPs (of the three known) have measured values of $P$ and $\dot{P}$) are instead binary MSPs that switch between an accreting/high X-ray emission state and a radio pulsar state. No clear grouping appears for these binaries sub-samples (see Fig. \ref{figure9}), as the intrinsic pulsar in them are similar. In future work, we shall supplement intrinsic variables with other representing the companion, the orbital parameters, and the environment of binary pulsars to try to address these issues. 

The branches departing from the main trunk cover deviations that are better represented by the variability in one or a few of the magnitudes considered. The extremes of each branch are thus extreme pulsars of the population in a particular way. 

For instance, the longest rightwards branch (moving along it towards the rightmost node) groups pulsars with increasing age, period, and period derivative. Pulsars in this branch are not particularly energetic nor do they have large magnetic fields at the light cylinder. Instead, they have an increasing surface field, reaching up to extreme values. The second-longest rightwards branch has similar behavior, but is formed by less magnetized objects at their surface, and also less energetic, slower, and older. Not surprisingly, then, magnetars are located in both of these branches, as is also the only XDIN, J1856-3754, quoted in the ATNF catalog. We show this in more detail in Fig. \ref{figure9}. Interestingly, the XDIN and the magnetars do not share the same branch. 

Some of the low-field magnetars, since they are more energetic, less magnetized objects that have nevertheless shown magnetar-flaring behavior, appear quite separate from the rest. In fact, the low-field magnetars depicted in the MST having $P<1$ are J1846-0258 \citep{Gavriil2008} and J1119-6127 \citep{Archibald2016}, and they appear in Fig. \ref{figure9} in the lower leftmost branch, one almost on top of the other in this scale. This is a very different location from where J2301+5852, J1647-4552, J1822-1604 and J0418+5732, the other low-field magnetars, are. They are all located at the end of the branch going rightwards, above the magnetars. 

The different branches at the bottom of the MST contain all energetic pulsars. They share the same values of light cylinder magnetic fields of the millisecond pulsars, at the other end of the MST, despite they are very different in almost every other aspect. The small branches in which the energetic pulsars divide at the bottom of the plot also separate them into those having a significantly higher value of different variables; like $\dot E$, $B_{lc}$, $B_s$. Fig. \ref{figure9} consistently shows how the pulsars associated with TeV confirmed or candidate pulsar wind nebulae (PWNe, \cite{HESSPWNe}) are essentially all located in these branches. Only three TeV PWNe, B1742-30(1)/J1745-3040, J1858+020/J1857+0143 and CTA 1/J0007+7303 are somewhat outliers. The former two are TeV PWN candidates, the oldest and less energetic PWNe in the population (see Table 4 of \cite{HESSPWNe}). The MST by itself cannot judge the reality of the association proposed, but emphasizes how distinct these two are with respect to the rest. The case of CTA 1 has been studied in detail as a possible PWN in the reverberation phase, and/or having a higher magnetization \citep{Martin2016}. Its peculiarity has been already noted from a physical standpoint, although it is less of an outlier in comparison to the rest of the PWNe population (both in the MST and in comparing PWNe models). 

Another interesting example of the MST view on pulsars is to note where the {\it Fermi}-LAT detected gamma-ray pulsars that are part of the ATNF fall on it (see \cite{2fpc,GLAMCOG}). Fig. \ref{figure9} shows two panels to this effect, where the ranges they have for the light cylinder magnetic field and spin-down power are noted. The detected gamma-ray pulsars cluster at specific locations of the tree, most of it being empty of gamma-ray emission. This corresponds to the fact that gamma-ray pulsars have high values of light cylinder magnetic field, $B_{lc}>100$ G, with most having actually $B_{lc}>10^3$ G, and relatively high spin-down power. Some other magnitudes are less decisive, e.g., there are gamma-ray pulsars across the full range of $\eta_{GJ}$-values. Along the branches where the {\it Fermi}-LAT pulsars lie there might be a close sequence of detected and non-detected pulsars, despite the MST clearly showing the similarity of their intrinsic properties. This is a representation that extrinsic features such as distance, environment, or geometry play a role in {\it Fermi}-LAT detectability at an individual level.

The {\it Fermi}-LAT pulsar isolated in the central part of the MST is J2208+4056, the only one depicted with $B_{lc}<100$ G. This pulsar has been noted by \cite{Smith2019} as having a spin-down ($\sim 8 \times 10^{32}$ erg s$^{-1}$) about three times lower than the previously observed gamma-ray emission death-line. The outlier {\it Fermi}-LAT pulsar in the left part of the MST is J1231-5113 and has an even lower spin-down power ($\sim 5 \times 10^{32}$ erg s$^{-1}$). In comparison, the other magnitudes $B_{lc}, \tau, \eta_{GJ}$ are similar to the rest of the gamma-ray population. 

We also investigate the position in the MST of the 40 rotating radio transients (RRATS) with known $P$ and $\dot P$ appearing in the ATNF \citep{Malusare2022,RRATalog}. RRATS are pulsars showing extreme radio variability, as most of them are discovered through their single, isolated pulses. Only one of the 40 RRATs considered is located in the main trunk.
This pulsar is near the degree 3 node J1828-1336 that separates the main trunk in the two central branches. 

The MST can also be used to analyze any other pulsar population, for instance, where do the pulsar's known glitches, or the intermittent and nulling pulsars, group. The online tool provided with this work promotes this kind of analysis.
\subsubsection{The MST view of evolutionary tracks}
\begin{figure}
\centering
    \includegraphics[width=0.5\textwidth]{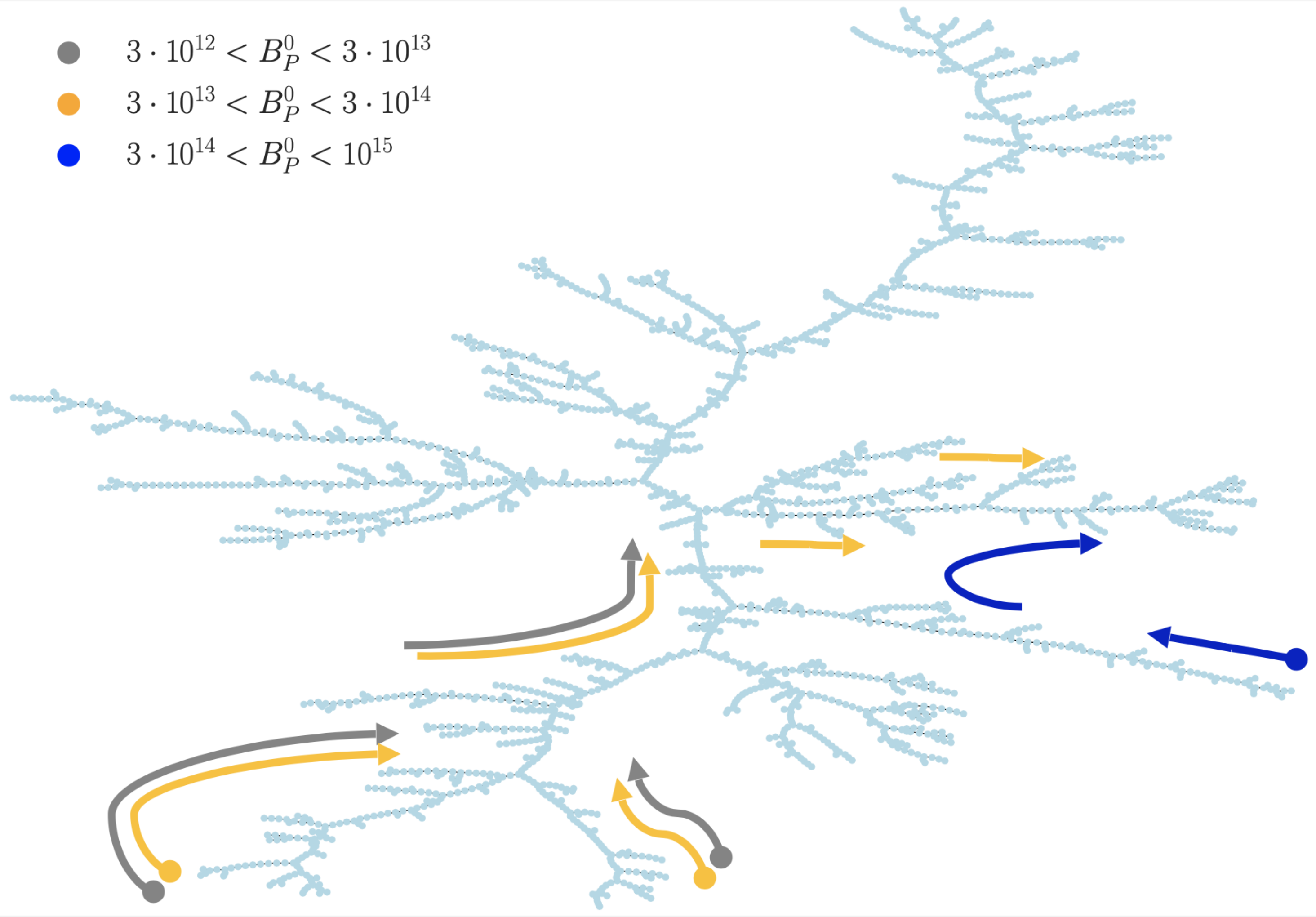}
    \caption{Representation of evolutionary tracks described in \protect\cite{Vigano2013} into the MST, according to their initial magnetic field.
}
    \label{figure10}
\end{figure}

While pulsars evolve, they change their timing parameters and move across the $P \dot P$-diagram. Evolutionary models are then constructed following the fully coupled evolution of temperature and magnetic field in neutron stars (e.g., see \cite{Vigano2013}). To simulate evolutionary tracks, we have created synthetic pulsars over the theoretical tracks of figure 10 of \cite{Vigano2013} and individually studied where would they fall in the MST should they be part of our sample. This is shown in Fig. \ref{figure10} where the arrows show increasing age at a fixed initial magnetic field and the rounded cap point in the origin of the arrows shows possible birthplaces. We find, in agreement with what was already discussed in Fig. \ref{figure6}, tracks are not randomly found in the MST.
There are two birth zones in it, at the bottom part where we find all energetic pulsars, and at the rightmost branch, where we find the magnetars. Then, for low initial magnetic field values, pulsars go on to die on the main trunk of the MST. When the field increases, the pulsars populate the middle branches. When the initial field is high enough to shift the birthplace from the energetic pulsar zone to the classical magnetar range, the evolution is mostly confined to the two rightmost branches in Fig. \ref{figure10}. In these branches then, we find pulsars evolving in the two directions (from the main trunk to the extreme and vice-versa) depending on their place of origin. 

\subsubsection{The MST as an alerting tool}
In addition of providing a descriptive perspective, the MST might be used as an alerting tool for pulsars-of-interest. These are some examples taking into account the MST location and the distance ranking. Implied connections are not always obvious using $(P, \dot P)$ only (see the discussion on the PCA and the distance ranking above and in the appendices). We use the web application provided with this work to note the following:
\begin{itemize}
\item Based on the location of the energetic low-field magnetars J1846-0258, J1119-6127 at the bottom part of the MST, and due as well to its nearness ranking, other pulsars with essentially the same characteristics are noted, in particular, J1208-6238. It has been suggested as a possible low-field magnetar in the literature (\cite{Clark2016}) and is second (first) in the distance ranking of J1846-0258 (J1119-6127) after J1119-6127 (followed by J1846-0258). PSRs J1513-5908 (in the composite SNR MSH 15-52), J1640-4631, and J1930+1852 follow in the ranking of J1846-0258; and J1640+4631, J1614-5048, and J1513-5908 do so in the ranking of J1119-6127.
\item Panel 5 of Fig. \ref{figure8} shows that few locations of the MST showing $B_{lc} \sim 100$ G or beyond and no detected gamma-ray pulsars yet. These regions become of special interest for future searches. In particular, those near J1231-5113 (which is already detected) in the MST appear to be promising potential targets.
A few neighboring pulsars to the outlier J1231-5113, at the end of this branch, also show a relatively large $B_{lc}$ with a similar range of spin-down power, and other variables, in comparison to {\it Fermi}-LAT pulsars. Likewise, PSR J1915+1616 and J2129+1210B at the end of the nearby branch are of interest. Again, note that both have $B_{lc}>10^3$ G, and $\dot E>10^{33}$ erg s$^{-1}$. 
\item The detection of the radio pulsar PSR J2208+4056 by the Fermi LAT in spite of its low spin-down power has been ascribed by \cite{Smith2019} to a possible case of favorable geometry. If this is the case, it may remain indeed isolated in the MST, where appears close to the main trunk. Its closest neighbours  (J0532-6639, J0502+4654, and J1848-0123) call for attention in order to test this.
\item Others pulsars-of-interest regarding their possible detection in gamma-rays maybe J1818-1607 and J1550-5418. These lie in the magnetar branch, where no other detected {\it Fermi}-LAT pulsar is located \citep{Li2017}. The latter authors established a {\it Fermi}-LAT integrated upper limit for J1550-5418 and  attempted folding, finding no signal. However, these pulsars have similar properties to other pulsars already detected by {\it Fermi}-LAT (ordered by distance, J1208-6238 occupies the third position, followed by  J1119-6217, both detected by {\it Fermi}-LAT).
In the case of the pulsar J1550-5418, among its ten closest pulsars according to the calculation of the Euclidean distance, the pulsars J1734-3333, J1746-2850 and J1726-3530 are in the sixth, eighth and ninth positions, respectively. They are located in the lower part of the MST, where there is a high density of {\it Fermi}-LAT pulsars, although they have not yet been detected in gamma-rays. The distance ranking of these two magnetars is uncommon to others: other magnetars in the branch have neither a {\it Fermi}-LAT pulsar nor other pulsars located in high-density areas of {\it Fermi}-LAT detections in the first positions of their distance ranking.
\item The pulsars J1842-0905 and J1457-5902, and J1413-6141 and J1907+0631 are the closest neighbours  to the pulsar wind nebulae J1745-3040/PWN B1742-30(1) and J0007+7303/PWN CTA 1, respectively. The latter are somewhat outliers of the pulsar wind nebulae population (see above) and thus their neighbours are of interest to test whether this region of the pulsar parameter space is prone to producing observable nebulae.
\item Finally, the higher-degree nodes --particularly those connected with the main trunk, the nodes that are extremes of significant branches, and possibly other topologically selected nodes, are suggested for individual study, as they may have undiscovered relevance. 
\end{itemize}
Note that we are not considering the physical distances to the pulsars in our MST, which is only based on intrinsic pulsar properties. Thus, it may well be the case (as it so happens)  that one of the noticed pulsars above is in the LMC (J0532-6639): in such occasions the MST may just be pointing that intrinsically this pulsar may emit similarly to others detected, despite it maybe too far to be seen with current instrumentation.

\section{Data availability and the pulsar tree web}
\label{pulsartreeweb}

The pulsar tree web\footnote{\url{http://www.pulsartree.ice.csic.es}}
accompanying this paper contains visualization tools and data to produce all plots and go beyond what has been presented in this paper. It allows the readers to gather information in terms of MST localization, $P\dot P$ comparison, and distance ranking. Among the functionalities included already, it can locate a given pulsar of the sample in the MST, $P \dot P$, $PC_1 PC_2$-diagrams and on any other diagram using the variables adopted in this paper; identify all properties of the given pulsar and all neighbouring nodes both in the MST; zoom around a given portion of the MST (or on any of the other diagrams); obtain tables of the properties of the nodes in the region of interest; obtain tables of the distance ranking for any pulsar, and more. As our research continues, we expect to upgrade the pulsar tree web with new functionalities.
\section{Discussion}
\label{discussion}

We have looked at the pulsar population from a different perspective from the usual $P\dot P$ diagram. Instead of considering just $P$ and $\dot P$ we used a set of 8 variables as proxies for intrinsic physical properties of all pulsars. While these variables all depend on $P$ and $\dot P$, the variance of the population is not fully contained by the variance of the latter quantities. $P$ and $\dot P$ are not the principal components. Distance ranking (or visualizations) based only on $P$ and $\dot P$ may hide interesting connections and mislead our intuition. We subsequently computed 
the pulsar MST, using a properly normalized Euclidean distance, discussed its properties and how the different classes of pulsars find an ordered place into it.

The MST approach offers applications beyond what we have described above. For instance, advanced analysis of the MST regarding clustering, centralization, betweenness, and closeness can illuminate physical connections and link pulsars among themselves. Similarly, a quantitative comparison of MSTs constructed for synthetic populations of pulsars offers a way to qualify the goodness of the population synthesis model that created them. The method used here can also be generalized to consider other variables in the distance definition and the PCA, and then in the MST done with them, allowing different problems to be treated. For instance, focus can be put on binary pulsars, and for them, environmental and orbital parameters (DM, orbital period, orbital size, nature of the companion, etc.) can be thought of as part of the distance definition. Another possibility in this regard would be to consider emission properties in particular regimes, distance definitions containing spectral parameters in a given energy range, or light curve properties. 
The forthcoming third {\it Fermi}-LAT pulsar catalog may be especially appropriate for working in this direction.
The MST technique usage in astrophysics is not as widespread yet as it is in other fields, albeit it has the potential to offer new perspectives in classification, clustering, source identification, and cross--correlation of source properties, including pulsars.

\section*{Acknowledgements}

We have made an extensive use of the ATNF Pulsar catalog v1.67: as updated on March 7th, 2022. This work has been supported by the grants PID2021-124581OB-I00, and PGC2018-095512-B-I00. C.R. is funded by the Ph.D. FPI fellowship PRE2019-090828, and acknowledges the graduate program of the Universitat Aut\`onoma of Barcelona. DFT acknowledges as well USTC and the Chinese Academy of Sciences Presidential Fellowship Initiative 2021VMA0001.  AP acknowledges partial support from a Ramon y Cajal fellowship RYC-2017-21810.
This work was also supported by the Spanish program Unidad de Excelencia María de Maeztu CEX2020-001058-M.

\section*{Data availability}

The data underlying this article are available in "The pulsar tree" web at the Institute of Space Sciences (ICE, CSIC) \url{http://pulsartree.ice.csic.es}

\bibliographystyle{mnras}
\bibliography{biblio}

\section*{Appendix 1: Principal Component Analysis
\label{sec:PCA_introduction}}

The principal component analysis (PCA) is a technique used to reduce the number of variables of a problem without having a significant loss of information, see, e.g.,  \cite{Shlens2014} for a review.
PCA is based on the search for existing correlations between the original variables involved, so as to find a new system to represent the data. The axes of this new system will be vectors that are linearly independent of each other, i.e. geometrically orthogonal, and oriented in the direction where they are able to encompass the largest possible variance of the processed data. This linear transformation is based on the eigenvector decomposition of the covariance matrix $C$ of the variables involved,

\begin{eqnarray}
C= \frac{1}{n-1}\sum _{i=1}^{n} (X_{i}-\hat{X})(X_{i}-\hat{X})^{T}
~;\label{eq:covariance.}
\end{eqnarray}
where $X$ is the data set in a matrix form, taking into account that $X\in \ R^{n\times N}$ and $C \in \ R^{N\times N}$, $n$ the size of the dataset, $N$ is the number of variables describing each member of the dataset, and $\hat{X}$ is the mean value of the variable. Furthermore, $C$ will by definition be a symmetric and positive semi-definite matrix which contains on its main diagonal the variance of the variables used. The eigenvector decomposition can then be done as,

\begin{eqnarray}
C\vec v = \lambda \vec v
~,\label{eq:eigendescomposition.}
\end{eqnarray}
where the eigenvectors $\vec v$ found from the matrix $C$ together with their eigenvalues $\lambda$, allow one to perform the above transformation from the original set of variables to the set of principal components (hereafter PCs). In this way, the eigenvectors $\vec v$ are the new axes on which the new dimensional space will be built, which have the intrinsic property of maximizing the variance of the treated data. Each $\vec v$ will have an associated $\lambda$ according to Eq. (\ref{eq:eigendescomposition.}). The eigenvalues serve to measure in which direction the data is most dispersed. Thus, it is useful to use the ratio of each $\lambda$ to the sum total of all eigenvalues to define what is known as explained variance. Said otherwise, the explained variance ratio is the percentage of variance that is associated with each of the PCs. The larger $\lambda$ is, the greater the explained variance covered by the resulting $\vec v$  will be. The cumulative explained variance is the sum of each explained variance. The new space may at most be $N-$dimensional, when all PCs have a non-null contribution to the variance of the population. This final set of PCs will be obtained by relating the original set of variables to the $\vec v$ that have been defined. This linear transformation follows 

\begin{eqnarray}
PC_{m} =\vec{v}_{m} X^{T}
\label{eq:PCs.}
\end{eqnarray}
where $X^{T}$ is the transpose matrix denoted in Eq. (\ref{eq:covariance.}). In this way, as $\vec v$ has dimension ($1\times N$) and $X^{T}$ has dimension ($N\times n$), the result is a row vector of size ($1\times n$) that will contain the $PC_m$ values for each member of the space.

\subsection*{Algebra with $PC_1$ and $PC_2$ }
\label{section:algebra}

The variables used depend on powers of $P$ and $\dot P$, and therefore once the logarithm is taken, the expressions defining each variable are linear. For example, $\log B_{s} = K + \log P + \log \dot{P}$, where $K$ is a constant (i.e. the equation of a plane, $Ax+By+Cz+D=0$). The eigenvalues associated with the principal components beyond $PC_1$ and $PC_2$ will be strictly null and therefore all the explained variance is accumulated in the first two PCs. 

Eqs. (\ref{eq:PC1_8V}) and (\ref{eq:PC2_8V}) are expressed as a function of the normalization of the logarithm of the variables, $(V_{l})^{\dag}$ where the sub-index $l$ indicates the logarithm of the generic variable $V$. Here, given that any of the variables we are considering can be written as $V(P,\dot {P}) = KP^{a} \dot P^{b}$, $V_{l} = \log V = K_{l} + aP_{l} + b\dot{P}_{l}$, the normalization is 

\begin{eqnarray}
        (V_{l})^{\dag} = \frac{V_{l} - Q_{2,V_{l}}}{IQR_{V_{l}}}=
        \frac{K_{l} + aP_{l} + b\dot{P}_{l} - Q_{2,V_{l}}}{IQR_{V_{l}}}.
    \label{eq:norm_log_var}
\end{eqnarray} 
Once Eq. (\ref{eq:norm_log_var}) is applied to all the variables, a summation will be obtained which leads to Eqs. (\ref{eq:PC1_2V}) and (\ref{eq:PC2_2V}). 

We now consider Fig. \ref{figure3}, where we show how a circle in the $P\dot P$-diagram would look in the principal components plane. The simulation of points located along a circle in the $P,\dot P$-diagram, defined by its logarithmic coordinates ($P_{l}, \dot P_{l}$), can be done by considering a fixed value $r^{2} = \Delta P_{l}^{2} + \Delta \dot{P}_{l}^{2}$, where $\Delta $ is used to represent that the circle can be displaced from the origin. Introducing Eqs. (\ref{eq:PC1_2V}) and (\ref{eq:PC2_2V}) into it, we get:

\begin{eqnarray}
r^{2} = 0.702\Delta PC_{1}^{2} + 0.148\Delta PC_{2}^{2} - 0.363 \Delta PC_{1}\Delta PC_{2}
~,\label{eq:ellipse}
\end{eqnarray}
which corresponds to an ellipse (i.e., $Ax^{2}+Bxy+Cy^{2}+Dx+Ey+F=0$) with an angle of rotation $ \cot(2\theta) = (A-C)/{B} \label{eq:angle_ellipse}$). Starting counter-clockwise from the positive axis of $PC_{1}$ this leads to  $\theta = 73.38^{\circ}$.

\section*{Appendix 2: Basics of graph theory}
\label{graph-principles}

\subsection*{Basic definitions for building a weighted graph}
\label{subsection:basics_graph}
The aim of graph theory is to establish relationships among objects based on the connections they have, ultimately representing them in a graph. $G(V,E)$ will denote a graph of a set $V$, of nodes $v$; and a set $E$, of edges $e$ joining these nodes. The latter are assigned with a weight $w$ (e.g., a distance value) representing the relationship between the nodes (see e.g.,\cite{Wilson2010}). These edges will not have any privileged addresses, and will therefore be treated as undirected connections. To assign the value of $w$ for a given edge, the simplest possibility is to assume the weight to be the Euclidean distance between the nodes in an $N-$dimensional space of interest (this is usually called an Euclidean graph),

\begin{eqnarray}
d_{nm}=\sqrt{ \sum_{j=1}^{N} \sum_{n=1}^{V} \sum_{m>n}^{V} (v_{jn}-v_{jm})^{2} }
~,\label{eq:d_eucl.}
\end{eqnarray}
where $d_{nm}$ represents the distance between variables that define each node individually.\footnote{As usual, a normalization process is adopted when there is the need to compare magnitudes with different units.} The total weight of a graph $G$ will be obtained from the sum of each specific weight on each edge,
\begin{eqnarray}
w(G)=\sum_{e\in E(G)} w(e).
\label{eq:weights}
\end{eqnarray}
Note that $e=\{v_n,v_m\}=\{v_m,v_n\}$. Moreover, $G$ has an edge between every pair of nodes, which makes it a complete (equivalent to fully connected) graph as well as undirected and weighted due to the edge characteristics described above.
A path is a sequence of nodes through $G$ in which no node is encountered more than once. Then $G$ is a connected graph where any pair of nodes can communicate through a path. If a path is closed it is called a cycle. In addition, the partition of $V$ into two sets ($S, V-S$) is called a cut. The set containing the edges $e=\{v_n,v_m\}$ whose $v_n \in$ S and $v_m \in$ V-S is defined as the cut set. The nodes of each graph can be classified according to the number of edges that connect to it, what is also known as the node degree $g(v)$. In turn, this value is the same as the number of adjacent nodes connected via incident edges to the node in question. The adjacency matrix is the backend of the graph. This matrix has dimensions ($|V| \times |V|$) where each $nm$-th element of this matrix indicates whether the nodes $e=\{v_n,v_m\}$ are adjacent. The value used for signalling the existence of adjacent nodes can either be 1 or $w(e)$ corresponding to the edge defined by both nodes. Note that the main diagonal of this matrix will be null as the graph $G$ does not contain loops (i.e., it does not contain cycles with just one node). The adjacency matrix can be used to compare two graphs quantitatively.

\subsection*{The minimum spanning tree (MST)}
\label{subsection:MST_graph}

$T(V, E')$ is a subgraph of $G(V,E)$, i.e., one whose nodes and edges come from the sets of nodes and edges of G, when $V\subseteq V$ --in our case the same-- and $E'\subseteq E$. Furthermore, we will say that $T$ is a spanning tree of $G$ when it does not contain cycles and connects all nodes of $G$. The following statements regarding $T$ are thus equivalent (see e.g., \cite{Tarjan1983})

 \begin{itemize}
     \item $T$ is a spanning tree of $G$.
     \item $T$ contains no cycles and is connected.
     \item $T$ is connected and $E'$ contains $|V|-1$ edges.
     \item $T$ has no cycles and $E'$ contains $|V|-1$ edges.
     \item $T$ is minimally connected: there is only one path to connect any two vertices of $T$, so by removing any edge $T$ it would be disconnected.
     \item $T$ is maximally acyclic: adding an edge to $T$ would form a cycle.
 \end{itemize}
The minimum spanning tree (MST) is a sub-graph of $G$ such that the sum of all weights, $\sum_{(e)\in T} w(e)$, is the smallest possible. 
In this way, the MST connects all the nodes $v$ taking into account the minimum distance between them at a local level, but under the condition that there is a global minimization for connecting the whole population of nodes. The solution of this kind of problem is obtain via greedy algorithms, which search the best possible solution in each iteration \citep{Roughgarden2019}. Using $G(V,E)$, such an algorithm joins $T \cup e_{i}$ (where $e_{i}$ as the edge with the smallest possible weight contained in $E$), at every step, unless the addition of this causes a cycle (see e.g., \cite{Wilson2010}). Here we use a special case of the greedy algorithm, called Kruskal's algorithm (see eg., \citep{Kruskal1956}) for getting an MST out of the complete $G$.To implement this algorithm, the following steps must be carried out

\begin{figure*}
  \includegraphics[width=1\textwidth]{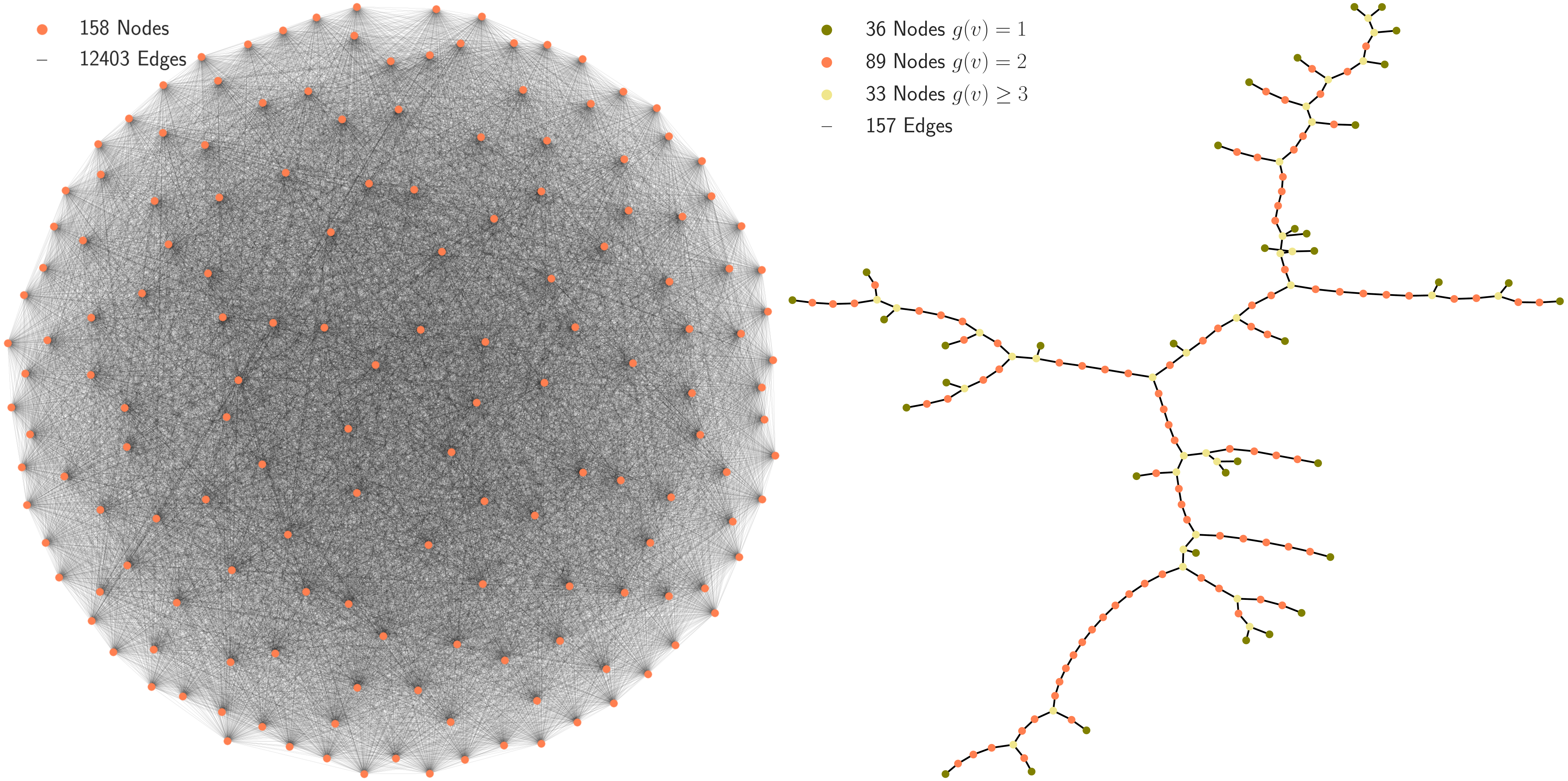}
  \centering
  \caption{Left: Complete, undirected and weighted graph $G(158,12403)$.
Right: Corresponding MST $T(158, 157)$, from the graph $G(158,12403)$ shown in the left. The nodes are noted according to the degree: $g(v)=1$ --olive, $g(v)=2$ --coral, and $g(v)\geq 3$ --yellow}.
  \label{figure11}
\end{figure*}

\begin{itemize}
  \item Order the edges by increasing weight.
  \item Choose the edge with the smallest $w(e)$ and add it to $E'$.
  \item Make sure that the chosen $e$ does not produce any cycle in the structure of $T$.
  \item The process finishes when all nodes $V$ are connected, thus resulting in a graph $T(V, E', w')$, where $|E'|=|V|-1$ and $w'(T)$ is the weight of $T$ according to Eq. (\ref{eq:weights})
\end{itemize}
The description of this algorithm would follow the pseudocode shown (see eg., \cite{Erickson2019}).
\begin{algorithm}
\renewcommand{\thealgorithm}{}
\caption{Kruskal $G(V,E,w)$}\label{pseudo_code}
\begin{algorithmic}[1]
\Require{sort $E$ in increasing order according to $w(e)$}
\State $T \gets (V, \varnothing)$
\For{each $v \in V$} 
\State \textit{MakeSet}($v$)
\EndFor
\For{i $\gets$ 1 to $|E|$}  
\State $e_{i}=\{v_{n}, v_{m}\}\ \gets e_{i}$ the lowest edge in $E$
\If{$Find(v_{n}) \neq Find(v_{m})$}
\State $Union(v_{n},v_{m})$
\State $T \gets T \cup e_{i}$
\EndIf
\EndFor
\State return $T$
\end{algorithmic}
\end{algorithm}
This description is supported by the implementation of the algorithm based on a union data structure, which operates on disjoint subsets of $V$, having the ability to support \textit{MakeSet}, \textit{Find} and \textit{Union} operations. Note that \textit{MakeSet} generates a subset for each node $v$. On the other hand, the \textit{Find} operation returns an identifier for the subset to which $v$ belongs. Finally, \textit{Union} decreases the number of subsets by merging those containing $v_{n}$ and $v_{m}$.
\\
To exemplify the structures of $G$ and $T$ as well as the complexity of working with a high number of nodes, Fig. \ref{figure11} shows a worked example of a complete, undirected and weighted graph $G$ with 158 nodes and 12403 edges (i.e., $158 \times 157 / 2$), using as weights the values of the Euclidean distances between nodes according to Eq. (\ref{eq:d_eucl.}). It can be seen that the number of edges is a quickly growing function of the number of nodes, and so are the computational time requirements. For example, the computational time to calculate the MST with  8 variables is about 288s. The right panel of Fig. \ref{figure11} shows $T$, the MST of $G$,  where we found only 157 edges in addition to the 158 nodes. 

\subsection*{MST properties}
Below we list properties associated with the MST (proofs can be found in graph theory books, e.g., \cite{Tarjan1983,Kleinberg2005,Erickson2019}). Considering that $T(V,E')$ is the MST of $G(V,E)$, the main properties are:
\begin{itemize}
\item Uniqueness:
If all $w(e)$ of $E(G)$ are distinct values, the resulting MST will be unique. This implies that the choice of any other greedy algorithm instead of Kuskal's as a solution to search for the MST (e.g. Prim's algorithm \cite{Prim1957}) would give the same results.
\item Cycle property:
From the construction of $T$ itself, any edge $e\in E$ such that $E' \cup \{e\}$ creates a cycle $C$ in $T$. On the other hand, if for some $e_{c}\in C$ it is determined that T($V, E'\cup\{e\}-\{e_{c}\}$) is a spanning tree. Thus, if $T$ is an MST then $w(e_{c})>w(e)$, otherwise, the edge with the largest weight contained in $C$ will not be contained in $E'$ for $T$ to remain an MST. Therefore, every node $v\in T$ has among its incident edges the $e$ with the smallest value $w(e)$. \item Cut property:
From the construction of $T$ itself, any edge $e\in E'$ such that $T(V, E'-\{e\}$) will cause a cut in $T$ separating it into 2 connected components. For the resulting cut set, denoted as $D$, it is observed that for any $e_{D}\in D$ such that $T(V, E'-\{e\}\cup\{e_{D}\})$ is a spanning tree. Thus, if $T$ is an MST then $w(e_{D})>w(e)$, otherwise, the lightest edge contained in $D$ must be contained in $E'$ for $T$ to remain an MST.
 \end{itemize}
\subsection*{Nearness in the context of different distances}
\label{nearness-comment}

\begin{figure}
\centering
  \includegraphics[width=0.4\textwidth]{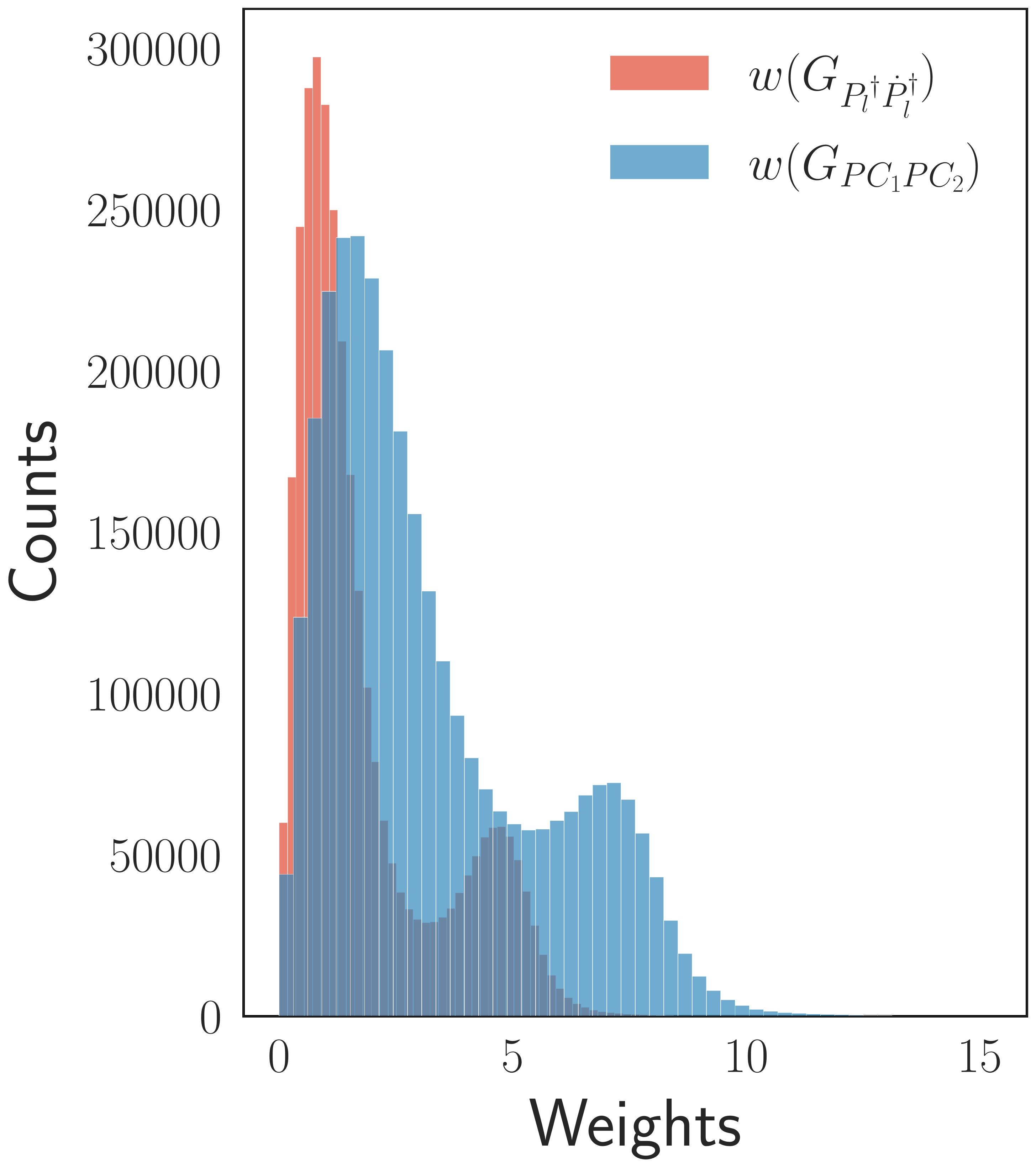}
  \caption{Distributions of the weights associated with the complete graph for two different distance definitions. We show in red distribution of distances based on the logarithms of the variables $P$ and $\dot P$ only --once they have been normalized; and in blue, the distribution of distances based on $PC_1$ and $PC_2$. 
}
  \label{figure12}
\end{figure}

Nearness between two pulsars depends on the definition of distance, and of the variables considered to compute it, thus the more complete this distance is, the better. One way to see this is by comparing the distributions of the weights associated to a complete, undirected and weighted graph using only $P,\dot P$ with those obtained with the full set of 8 magnitudes considered in the text (equivalently, their two principal components). This is shown in Fig. \ref{figure12}, and relates to the discussion of the bottom panels of  Fig. \ref{figure3}. The distributions are different. In addition, just from the full set of weights $w(G)$ it is possible to know which would be the nearest neighbours of any randomly chosen pulsar. It can be seen that neighbouring pulsars differ according to each distance definition. Considering the nearest three neighbours for every pulsar using a distance based on $P,\dot P$ and comparing with the ones obtained using the full set of variables of interest, we find that 45\% of the population incorporates a new pulsar even in the first three places of the ranking. The ranking positions, even when the same pulsars are concerned, may change in many cases: 40\% of the population has at most one pulsar in the same position.

\subsection*{Visual nearness in the MST and ranking}
\label{subsection:ranking_MST}

Due to MST properties, we know that the first neighbour in the distance ranking from a given source is indeed one of the attached nodes in the MST. However, the second neighbour and beyond in the distance ranking do not necessarily reflect vis-a-vis the visual appearance of the MST. Thus, visual nearness from a given source in the MST is not a one-to-one overall translation of the distance ranking from that source. The MST does not minimize the distance to a given node only, but the total distance needed to visit the whole population. This introduces a global perspective in the selection of how particular sources are connected to the rest, which is what ultimately serves for clustering analysis techniques. 

\subsection*{Representing the MST: how rendering is done}
\label{subsection:representing_MST}

Given that an MST lacks a fixed set of axes, the rendering of the MST (angles among branches, orientation, branch direction) does not hold any particular meaning, only the connections among nodes do. A different representation could be chosen conserving the same properties (e.g., each node being linked to the same neighbours, being of the same order, preserving the same sequence of all variables along all branches, etc.). The nearness and connections described in one MST would be the same as in the other, despite the overall appearance differing. An obvious example is to take one of the branches that go rightwards in our MST and force it to go leftwards. This would produce exactly the same MST (the same adjacency matrix, in technical terms), and everything we have said looking to the MST with the same branch pointing rightwards would apply for the new rendering as well. It is not the appearance of the overall graph what is important, but the graph properties. In our representation we are using the {\sc neato} program that uses the Graphviz python library \citep{graphviz}, which searches the minimization of a pseudo-energy to select the orientation of the different components. We have verified using earlier versions of the ATNF catalog how the global appearance of the MST changes whenever a significant number of nodes are added (usually, the rendering does not change when adding a few nodes). Even when the visual appearance of the tree may differ, all similarity properties (e.g., the relative localization of members of sub-populations) and physical connections described are maintained.

\subsection*{Code implementation}

Our code is built on python {v 3.10.4} \citep{python}, in which we implement the Psrqpy package \citep{psrqpy}  to deal with the population of pulsars from the ATNF catalog. To create the graphs we use the NetworkX \citep{networkx} and the Graphviz \citep{graphviz} libraries. On the other hand, the SciPy library \citep{scipy} contains Kruskal's algorithm according to which we have been able to calculate the MST. For the application of the PCA, we used the Scikit-learn library \citep{scikit}. The previous libraries and packages contain as requirements other well-known libraries such as NumPy \citep{numpy}, Pandas \citep{pandas} and Matplotlib \citep{matplotlib} through which, in addition, we have been able to develop those parts of the code that were necessary to obtain the results seen. The app is done using Bokeh \citep{Bokeh}.

\bsp	
\label{lastpage}
\end{document}